\documentclass[twocolumn]{aastex631}
\newcommand{\rev}[1]{#1}

\begin{document}

\title{
Rapid formation of Gas Giant Planets via Collisional Coagulation from Dust Grains to Planetary Cores
}

\shorttitle{
Collisional Growth from Dust Grains to Planets
}

\email{hkobayas@nagoya-u.jp}

\author{Hiroshi Kobayashi}

\affiliation{Department of Physics, Nagoya University, Nagoya, Aichi 464-8602, Japan}

\author{Hidekazu Tanaka}
\affiliation{Astronomical Institute, Tohoku University, Aramaki, Aoba-ku, Sendai 980-8578, Japan}

\begin{abstract}
Gas-giant planets, such as Jupiter, Saturn and massive exoplanets, were
formed via the gas accretion onto the solid cores each with a mass of
roughly ten Earth masses. 
%\citep{ikoma00}. 
However, rapid radial migration due to disk-planet interaction prevents the formation of such massive cores via
planetesimal accretion. 
%\citep{tanaka,kobayashi11}. 
Comparably rapid core growth via pebble accretion requires very massive protoplanetary disks because most pebbles fall into the central star. 
%\citep{lambrechts14}.
Although planetesimal formation, 
planetary migration, and gas-giant core formation have been studied with
much effort, the full evolution path from dust to planets are still
uncertain. 
Here we report the result of full simulations for collisional
evolution from dust to planets in a whole disk.  
Dust growth with realistic porosity allows the formation of icy planetesimals in the inner disk ($\la 10$\,au), while pebbles formed in the outer disk drift to the inner disk and there grow to planetesimals.
The growth of those pebbles to planetesimals suppresses their radial drift 
and supplies small planetesimals sustainably in the vicinity of cores. 
This enables rapid formation of sufficiently massive planetary cores within 
0.2-0.4 million years, prior to the planetary migration. 
Our models shows first gas giants form at 2-7\,au in rather common protoplanetary disks, in agreement with the exoplanet and solar systems.
\end{abstract}

\section{Introduction}

Gas giant planets are formed via the rapid gas accretion of solid cores each with about $10\,M_\oplus$ in protoplanetary disks \citep{ikoma00}, where 
$M_\oplus$ is the Earth mass. 
The formation of cores via the accretion of 10 km sized planetesimals is in the Jupiter-Saturn forming region estimated to be $ \sim 10^7$ years \citep{kobayashi11}, which is longer than the disk lifetime 
\citep[several million years,][]{haisch01}. 
In addition, the cores undergo the fast migration
caused by the tidal interaction with the disk 
 \citep[called ``Type I'' migration,][]{ward97}. 
They are lost prior to the gas accretion if the core
formation timescale is longer than the migration timescale 
\citep[$\sim 10^5$years,][]{tanaka}.
Recently, the rapid accretion of submeter-sized bodies (called ``pebbles'' in the context of planet formation) is argued  \citep{ormel10b}. Pebbles form via collisional coagulation in the outer disk and then drift to the core-growing inner disk.  
The accretion of such bodies may lead to the formation of massive cores in a timescale ($\sim 10^5$ years) comparable to the migration timescale \citep{lambrechts14}. However, this process requires a massive disk because the pebble accretion is lossy. 
The capture rate of pebbles by a single planetary core is evaluated to be below 10\% \citep{ormel10b, lin18, okamura21}. % \citep{ormel10b}. 
Hence the total pebble mass of a few hundred Earth masses is required for the formation of a core 
with $10 M_\oplus$ (see more detailed estimate in \S \ref{sc:mdisk}), while protoplanetary disks with such a large solid mass
are very rare \citep{mulders21}. 

As a process achieving a high conversion rate from dust or pebbles 
to kilometer-sized or larger bodies, planetesimal formation via collisional growth of icy pebbles is one of the most probable candidates.
Recent models for collisional evolution of dust grains showed that pebbles grow to planetesimals in inner disks ($\lesssim 10$\,au) in the realistic bulk density evolution model \citep{okuzumi12}. This process enhances the solid surface density in the inner disk, while 10 km sized or larger planetesimals slowly accrete onto cores. If the accretion of (sub-)kilometer sized planetesimals effectively occurs prior to planetesimal growth, cores are expected to grow in a short timescale ($\sim 10^4$ years). 
In order to confirm rapid core formation, 
the treatment fully from dust to cores in a whole disk is required. 

In this paper, we investigate the formation of solid cores of giant
planets from dust grains in protoplanetary disks. In
\S~\ref{sc:disk_model}, we introduce the disk model that we apply. In
\S~\ref{sc:estimate}, we analytically estimate the growth timescales of
solid cores via planetesimal and pebble accretion, respectively. In
addition, we also estimate the minimum disk masses required for the formation of single gas-giant cores via pebble accretion. In \S~\ref{sc:DTPSmodel}, we model
a simulation for the collisional evolution of bodies from dust to planet
(“Dust-to-planet” simulation; Hereafter, DTPS), taking into account the bulk density evolution of dust aggregates.  
This model consistently includes planetesimal and pebble accretion. 
In
\S~\ref{sc:result}, we show the result of a DTPS, where the rapid formation of solid cores via the accretion of planetesimals formed via drifting pebbles. In
\S~\ref{sc:discussion}, we discuss the locations of giant planets in the solar system or for exoplanets, based on the results of DTPSs. 
In \S~\ref{sc:summary}, we summarize our findings.

\section{Disk Model}
\label{sc:disk_model}

Planet formation occurs in a protoplanetary disk. 
We consider a power-law disk model for gas and solid surface densities, $\Sigma_{\rm g}$ and $\Sigma_{\rm s}$, as 
\begin{eqnarray}
 \Sigma_{\rm g} &=& \Sigma_{\rm g,1} (r/1\,{\rm au})^{-1},\label{eq:sigmag} \\
 \Sigma_{\rm s} &=& \Sigma_{\rm s,1} (r/1\,{\rm au})^{-1},
\end{eqnarray}
where $\Sigma_{\rm g,1} = 480 \, {\rm g/cm}^2$ and $\Sigma_{\rm s,1} = 8.5 \,{\rm g/cm}^2$ are the gas and solid
surface densities at $1\,{\rm au}$, respectively, and $r$ is the distance from the host
star. The given solid/gas radio is the same as that in the
minimum mass solar nebula (MMSN) model beyond
the snowline  \citep{hayashi81}. However, we apply the shallower power law
index than the the MMSN model 
according to the observation of protoplanetary disks  \citep{andrews07}. 
The surface densities $\Sigma_{\rm g}$ and $\Sigma_{\rm s}$ at $r = 12.5\,{\rm au}$ correspond to those in the
MMSN model, while $\Sigma_{\rm g}$ and $\Sigma_{\rm s}$ are smaller than
those in the MMSN in the inner disk. 
The typical sizes of observed disks are $\approx 100\,{\rm au}$ \citep{andrews10}. 
We set disk radii $\approx 108$\,au: Disk masses correspond to $0.037 M_\sun$ ($\approx 220 M_\oplus$ in solid). 

We set the temperature at the disk midplane as 
\begin{equation}
 T = 200 \left(\frac{r}{1\,{\rm au}}\right)^{-1/2} \,{\rm K}. 
\end{equation}
The radial dependence of temperature is the same as the MMSN. However,
we apply a low temperature according to \citet{brauer08} 
because of optically thick disks. 
%the luminosity of a protostar is
%smaller than that of the main-sequence star 

\section{Analytic Estimate}
\label{sc:estimate}

\subsection{Core-Growth and Migration Timescales}
\label{sc:timescale}

We here estimate the growth timescale of a solid core growing via collisions with planetesimals. Taking into account the gravitational focusing, the growth rate is given by 
$dM_{\rm p}/dt \sim  2 \pi G M_{\rm p} R_{\rm p} \Sigma_{\rm s} \Omega / v_{\rm rel}^2$ \citep[e.g.,][]{goldreich04}, where $M_{\rm p}$ and $R_{\rm p}$ are the mass and radius of a planetary embryo, respectively, $\Omega$ is the Keplerian frequency, $v_{\rm rel}$ is the relative velocity between the core and planetesimals, and $t$ is the time. 
For a solid core with $M_{\rm p} \sim 10 M_\oplus$, planetary atmosphere enhances the collisional radius of the planet. The growth rate is estimated using $R_{\rm e}$ instead of $R_{\rm p}$, where $R_{\rm e}$ is the enhancement radius via atmosphere  \citep{inaba_ikoma03}. Assuming the relative velocity is determined by the equilibrium between gas drag and the stirring by the core, we estimate the growth timescale 
$t_{\rm grow} = M_{\rm p} / \dot M_{\rm p}$
 \citep[e.g.,][]{kobayashi+10,kobayashi11}
\begin{eqnarray}
 t_{\rm grow} &\approx& 8.4 \times 10^6 \left(\frac{\Sigma_{\rm s}}{1.2\,{\rm g\,cm}^{-2}}\right)^{-1} 
\left(\frac{R_{\rm e}/R_{\rm p}}{3}\right)^{-1}
\left(\frac{m}{10^{19}\,{\rm g}}\right)^{2/15}
\nonumber
\\
&& \times 
\left(\frac{\rho_{\rm b}}{1.4\,{\rm g \, cm}^{-3}}\right)^{4/15}
\left(\frac{M_{\rm p}}{10M_\oplus}\right)^{-1/3}
\left(\frac{r}{7\,{\rm au}}\right)^{13/10}
\nonumber
\\
&& \times 
\left(\frac{T}{76\,{\rm K}}\right)^{1/5}
%\nonumber
%\\
%&& \times 
\left(\frac{\Sigma_{\rm g}}{69\,{\rm g\, cm}^{-2}}\right)^{-2/5}
\,{\rm yr}, 
\label{eq:tg} 
\end{eqnarray}
where $m$ and $\rho_{\rm b}$ are the mass and bulk density of planetesimals, respectively, and $R_{\rm e} = 3 R_{\rm p}$ is used from the previous estimate  \citep[see Figure 1 in ][]{kobayashi11} and the values related to the disk are chosen from those of the given disk at $r = 7$\,au. Therefore, the growth timescale is comparable to or longer than the lifetimes of protoplanetary disks $\sim 10^6$ years. 

The gravitational interaction between a solid core and the protoplanetary disk induces radial migration of the core. The orbital decay timescale for the type I migration is estimated to be 
 \citep[e.g.,][]{tanaka}
\begin{eqnarray}
 t_{\rm mig} &=& \Gamma^{-1} \left(\frac{\Sigma_{\rm g} r^2}{M_*}\right)^{-1} 
  \left(\frac{M_{\rm p}}{M_*}\right)^{-1} \left(\frac{h_{\rm g}}{r}\right)^{2} \Omega^{-1}, 
\nonumber
  \\
&=& 1.6 \times 10^5 \left(\frac{\Gamma}{4}\right)^{-1} 
\left(\frac{M_{\rm p}}{10 M_\oplus}\right)^{-1}
\left(\frac{\Sigma_{\rm g}}{69\,{\rm g \,cm}^{-2}}\right)^{-1}
\nonumber
\\
&& \times 
\left(\frac{h_{\rm g}}{0.05a}\right)^2
\left(\frac{r}{7\,{\rm au}}\right)^{-1/2}
\left(\frac{M_*}{M_\odot}\right)^{3/2} {\rm yr}, \label{eq:mig1} 
\end{eqnarray}
where $\Gamma$ is the dimensionless migration coefficient, $h_{\rm g}$ is the scale height of the disk, the values of $\Sigma_{\rm g}$ and $h_{\rm g}$ are chosen from the given disk at $7\,$au. In the isothermal disk, $\Gamma \approx 4$  \citep{tanaka}. The formation of planets prior to the orbital decay requires $t_{\rm grow} \ll t_{\rm mig}$. 
Once planetary embryos reaches $\sim 10 M_\oplus$, the rapid gas accretion of planetary embryos occurs  \citep[e.g.,][]{mizuno80}. 
Gas giant planets formed by the gas accretion open up the gap around their orbits and the migration timescale is then much longer than the estimate in Eq.~(\ref{eq:mig1}) because of the onset of type II migration.  
%\citep[e.g.,][]{ida_lin08}. 
Therefore, the formation timescale of a massive core with $10 M_\oplus$ is required to be comparable to or shorter than the type I migration timescale. 

% The accretion of smaller planetesimals shortens $t_{\rm g}$. The enhancement $R_{\rm e}/R$ is the dependence as $R_{\rm e}/R \propto m^{-1/9} \rho_{\rm b}^{-2/9}$. The bulk density of planetesimals increases with $m$ due to self-gravity, resulting in $\rho_{\rm b} \approx 0.3 (m/10^{19})^{2/5} \,{\rm g\,cm}^{-3}$  \citep{kataoka13}. 
% Therefore, the growth timescale for $M_{\rm p} = 10 M_\oplus$ is estimated to be 
% \begin{equation}
%  t_{\rm grow} \approx 
% 1.9 \times 10^5 \left(\frac{\Sigma_{\rm s}}{1.2\,{\rm g\,cm}^{-2}}\right)^{-1} 
% \left(\frac{m}{10^{16}\,{\rm g}}\right)^{11/25}
% \left(\frac{a}{7\,{\rm au}}\right)^{13/10} \,{\rm yr}.\label{eq:tg2} 
% \end{equation}
% This timescale is comparable to $t_{\rm mig}$. In addition, the collisional growth of dust grains produce planetesimals in 10\,au, which enhances the solid surface density  \citep{okuzumi12}. 

The collisional growth of dust grains forms pebbles, which drift inward. 
\rev{
The growth rate of a planetary core via pebble accretion is given by 
\begin{equation}
 \frac{dM_{\rm p}}{dt} = \varepsilon \frac{d M_{\rm F}}{dt},\label{eq:pebble} 
\end{equation}
where $d M_{\rm F} / dt$ is the mass flux of pebbles across the orbit of the core and $\varepsilon$ is the accretion efficiency of drifting pebbles. 
From $d M_{\rm F} /dt $ given by Eq.~(14) of \citet{lambrechts14}, }
the core growth timescale via the accretion of drifting pebbles 
is estimated to be  
\begin{eqnarray}
 t_{\rm grow,pe} &=& 2.0 \times 10^5 \left(\frac{\varepsilon}{0.1}\right)^{-1}
  \left(\frac{\Sigma_{\rm g,1}}{480\,{\rm g\, cm}^{-2}}\right)^{2/5}
\nonumber
\\
&& \times 
  \left(\frac{\Sigma_{\rm s,1}}{8.5\,{\rm g\, cm}}\right)^{-5/3}
  \left(\frac{t}{10^5{\rm yr}}\right)^{-1/3} \,{\rm yr},\label{eq:tgpe} 
\end{eqnarray}
where 
the value of $\varepsilon$ 
is used for typical pebble-sized bodies \citep{ormel10b,okamura21}. 
%Although the pebble accretion requires a massive core prior to the significant pebble flux, 
Although the collisional cross sections for the pebble accretion are large thanks to strong gas drag for pebbles, 
the solid surface density of drifting pebbles is much lower because to their rapid drift. 
The growth timescale $t_{\rm grow,pe}$ is therefore comparable to the migration timescale.  

\subsection{Total Mass Required for Pebble Accretion}
\label{sc:mdisk}

\rev{
The required mass for the formation of a core with $M_{\rm p}$ via pebble accretion is given by $M_{\rm F}$ obtained from the integration of Eq.~(\ref{eq:pebble}). }
Integrating Eq.~(\ref{eq:pebble}) with the relation $\varepsilon \propto M_{\rm p}^{2/3}$ \citep{ormel10b, okamura21}, we have 
\begin{equation}
 M_{\rm F} = 3  M_{\rm p} / \varepsilon(M_{\rm p})
,\label{eq:pebble2} 
\end{equation}
where we assume the initial core mass is much smaller than $M_{\rm p}$. 

\begin{figure}[htb]
\plotone{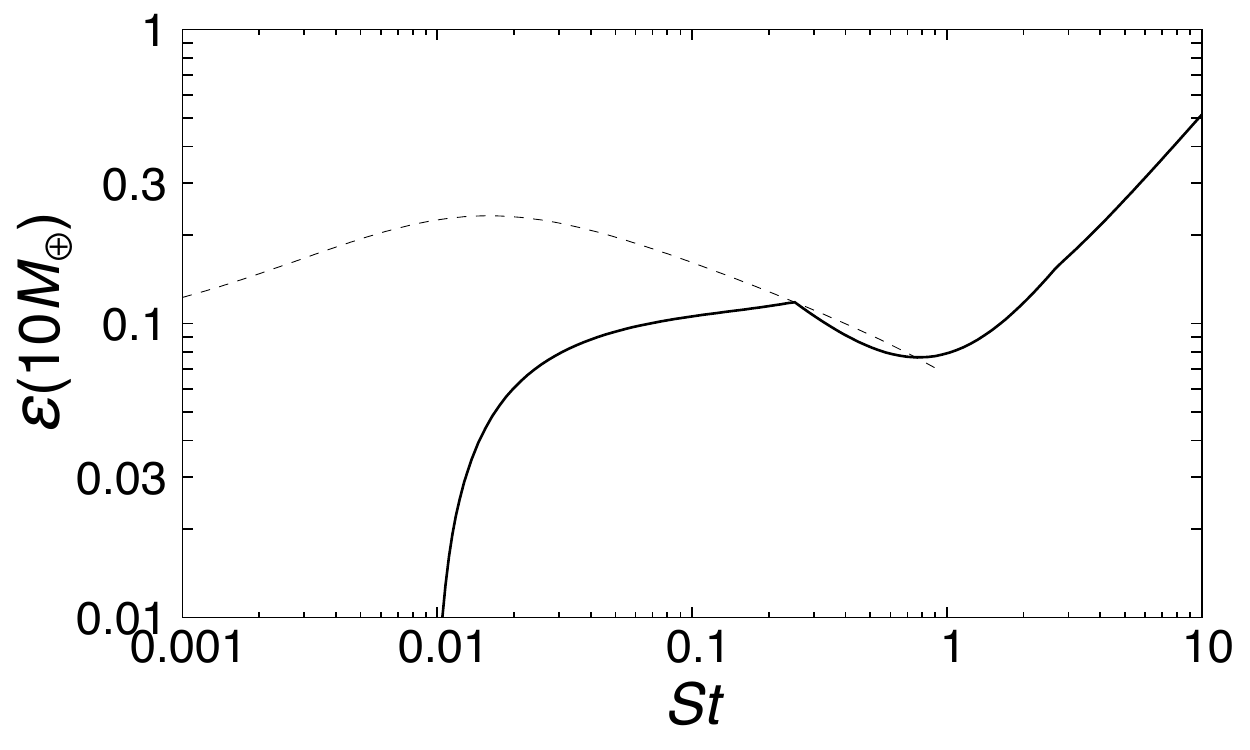}
%\plotone{epsi.eps}
\caption{Accretion efficiency of a $10\,M_\oplus$ core, $\varepsilon(10\,M_\oplus)$ at 7\,au in the disk given in \S~\ref{sc:disk_model} and \ref{sc:method}, as a function of the dimensionless stopping time $St$. Solid curve indicates $\varepsilon$ derived based on the gas densities and velocities obtained from hydrodynamic simulations \citep{okamura21}. For reference, the dotted curve shows $\varepsilon$ derived under the assumption of unperturbed circular gas motions \citep{ormel_liu18}. 
\label{fig:epsi}
} 
\end{figure}

The pebble mass required for core formation, $M_{\rm F}$, is inversely proportional to $\varepsilon(M_{\rm p})$ (see Eq.~\ref{eq:pebble2}). 
Figure \ref{fig:epsi} shows $\varepsilon(10\,M_\oplus)$ as a function of the dimensionless stopping time due to gas drag, $St$. 
For $St \approx 0.2$--1, $\varepsilon$ increases with decreasing $St$, because of slow drift for low $St$. However, for $St \la 0.2$, $\varepsilon$ decreases with decreasing $St$. The horseshoe flow reduces the accretion band of pebbles for $St \approx 0.02$--0.2, while the outflow around the Bondi sphere disturbs the accretion of pebbles \citep{kuwahara19, okamura21}. These effects reduce $\varepsilon$ significantly. 
For $M_{\rm p}\sim 10 M_\oplus$, $\varepsilon$ is estimated to be $0.1$ or smaller for pebble-sized bodies (Figure \ref{fig:epsi}). It should be noted that the estimate of $\varepsilon$ ignoring the realistic gas flow around a planet gives fatal overestimates for $St \la 10^{-2}$ \citep[see Figure~\ref{fig:epsi} and compare the formulae by][]{okamura21,ormel_liu18}. 

Although $\varepsilon \ll 0.1$ for $St \ll 0.1$, we consider $\varepsilon=0.1$ for $St \approx 0.1$. Then the pebble mass required for core formation, $M_{\rm F}$, is estimated to be about $300 (\varepsilon /0.1)^{-1} M_\oplus$ from Eq.~(\ref{eq:pebble2}). 
The pebble mass $M_{\rm F}$ is limited by the total solid mass in a disk and 
thus the minimum disk mass required for a single core formation is estimated as 
%The maximum value of $M_{\rm F}$ corresponds to the solid mass in a disk. 
%According to Eq.~(\ref{eq:pebble2}), the disk mass required for a single core formation is estimated to be 
\begin{equation}
 M_{\rm disk,min} = 0.05 \left(\frac{\varepsilon}{0.1}\right)^{-1} \left(\frac{M_{\rm p}}{10M_\oplus}\right) \left(\frac{\Sigma_{\rm d,1}/\Sigma_{\rm g,1}}{0.018}\right)^{-1}
M_\odot,\label{m_disk_req}
\end{equation}
where $\Sigma_{\rm s,1}/\Sigma_{\rm g,1}$ is the metallicity in the disk. However, most protoplanetary disks are less massive than $0.1 M_\sun$ \citep{andrews10} and disks with the solid mass of $300 M_\oplus$ are very rare
even in Class 0 objects \citep[$\sim 1$\%,][]{mulders21}. 
Note that the required solid mass of $300\,M_\oplus$ for pebble accretion
is the minimum value. If smaller pebbles with $St \ll 0.1$ are considered, 
the required solid mass is much more than $300 \, M_\oplus$.
Therefore, it seems difficult to explain giant exoplanets existing rather
commonly \citep[$\sim 10\%$,][]{mayor11} with pebble accretion. 
%Therefore, 
%it seems difficult to explain giant exoplanets existing 
%rather commonly with pebble accretion.
%formation of gas giant planets in the common disks ($\sim 0.03 M_{\sun}$) is difficult. 
To reconcile this issue, we needs to increase $\varepsilon$ 
due to collisional growth of drifting pebbles \citep[see $\varepsilon$ for $St \gg 1$ in Figure ~\ref{fig:epsi} and the discussion by][]{okamura21}.
%\citep{okamura21}. 

The planetesimal formation from icy pebbles would be 
a possible process, which achieve a high conversion rate 
from pebbles to kilometer-sized or larger bodies.
To consider the collisional growth of pebbles into planetesimals, 
we need to review collisional fragmentation. 
The collisional simulations of icy dust aggregates shows 
the fragmentation velocity of aggregates, $v_{\rm f}$, depends on the interaction of monomers determined by the surface energy of ice $\gamma_{\rm ice}$, given by $v_{\rm f} = 80 \, (\gamma_{\rm ice} / 0.1 \,{\rm J \, m}^{-2})^{5/6}\,{\rm m \, s}^{-1}$ for aggregates composed of sub-micron sized monomers \citep{wada13}. The collisional velocities for pebble-sized bodies are mainly smaller than 50\,${\rm m \, s}^{-1}$, so that collisional fragmentation is negligible if $\gamma_{\rm ice} \sim 0.1 \, {\rm J \,m}^{-2}$. 
The surface energy of ice was estimated to be much lower than $0.1 \, {\rm J \, m}^{-2}$ from the measurement of the rolling friction force between 1.1 millimeter sized particles in laboratory experiments \citep{musiolik19}. 
However, the distinction between rolling and slide forces is difficult for such large particles so that 
\citet{kimura20a} explained 
the measurements including the temperature dependence 
as the slide forces given by the tribology theory with quasi-liquid layers without low $\gamma_{\rm ice}$. 
In addition, the measurement in laboratory experiments showed 
the tensile strength of aggregates for ice is comparable to that for silicates, implying that $\gamma_{\rm ice}$ is as small as the silicate surface energy, $\gamma_{\rm sil}$ \citep{gundlach19}. 
\citet{kimura20b} explained the measured tensile strengths for ice and silicate by the Griffith theory using $\gamma_{\rm ice} \sim \gamma_{\rm sil} \sim 0.1 \,{\rm J \, m}^{-2}$. From a physical point of view, the surface energy should be grater than the surface tension, which is $\approx 0.08 \, {\rm J \, m}^{-2}$ even in the room temperature. Therefore, collisional fragmentation is negligible for pebble growth. 

We additionally discuss the effect of collisions with large $m_1/m_2$, where $m_1$ and $m_2$ are the masses of colliding bodies ($m_1 > m_2$). Erosive collisions, large $m_1/m_2$ collisions with velocities higher than $v_{\rm f}$, reduce the masses of the larger colliding bodies, which inhibit the growth via collisions with large $m_1/m_2$. \citet{krijt15} claimed the growth of pebbles were stalled by erosive collisions under the assumption that $v_{\rm f}$ for $m_1/m_2 > 100$ is much smaller than that for $m_1\sim m_2$. However, this assumption is inconsistent with the impact simulations of dust aggregates. Recent impact simulations with $m_1 /m_2 \ga 100$ show $v_{\rm f}$ for larger $m_1 /m_2$ is higher than that for $m_1 \sim m_2$ \citep{hasegawa21}. Therefore, erosive collisions are insignificant for pebble growth.

\section{Models for DTPS}
\label{sc:DTPSmodel}

We develop a simulation for the collisional evolution of bodies from dust to planets (``Dust-to-planet'' simulation; Hereafter, DTPS). We here introduce the model for the DTPS. 

\subsection{Collisional Evolution}
\label{sc:method}

Collisions between bodies lead to planet formation. 
The surface number density $n_{\rm s}$ of bodies with mass $m$ at the distance $r$ from the  host star with mass $M_*$ evolves via collisions and radial drift. 
The governing equation is given by 
\begin{eqnarray}
\frac{\partial}{\partial t} n_{s}(m,r) &=& \frac{1}{2}
\int_0^\infty dm_1 \int_0^\infty dm_2 n(m_1,r) n(m_2,r) 
\nonumber\label{eq:coag} 
\\ && \times K(m_1,m_2) \delta(m-m_1-m_2)
\nonumber
\\ && 
- n(m,r) \int_0^\infty dm_2 n(m_2,r) K(m,m_2)
\nonumber
\\ && 
- \frac{1}{r} \frac{\partial}{\partial r} [r n(m,r) v_r], 
\end{eqnarray}
where %$t$ is the time, 
$K(m_1,m_2)$ is the collisional kernel between
bodies with masses $m_1$ and $m_2$ 
%$f(m,m_1,m_2)$ is the function determining the outcome of a collision between $m_1$ and $m_2$, 
and 
$v_r$ is the radial drift velocity. 
We adopt 
\begin{equation}
 v_{r} = v_{\rm drag} + v_{\rm mig}, 
\end{equation}
where $v_{\rm drag}$ and $v_{\rm mig}$ are the radial drift velocity due to gas drag and the type I migration. 
We model as \citep[see Appendix \ref{sc:app_drift} for $v_{\rm drag}$ and ][for $v_{\rm mig}$]{tanaka}
\begin{eqnarray}
 v_{\rm drag} &=& - \frac{2 r \Omega St}{1+St^2}
\left(
0.34 
%\left[
%\frac{E(3/4)+K(3/4)}{3\pi}
%\right]^2
e^2 + 
\frac{4 i^2 
}{\pi^2}
+ \eta^2
\right)^{1/2}, \\
 v_{\rm mig} &=& 
%- \frac{r}{t_{\rm mig}}, 
- \Gamma
\left(\frac{\Sigma_{\rm g} r^2}{M_*}\right)
  \left(\frac{m}{M_*}\right) \left(\frac{h_{\rm g}}{a}\right)^{2} r \Omega
\end{eqnarray}
where $\Gamma = 4$ is the dimensionless migration coefficient \citep{tanaka}, 
%$K$ and $E$ are the complete elliptic integrals of the first and second kinds, 
$e$ and $i$ are the orbital eccentricity and inclination, $St$ is the dimensionless stopping time due to gas drag, and 
\begin{eqnarray}
\eta &=& - \frac{1}{2} \left(\frac{c_{\rm s}}{r\Omega}\right)^2
 \frac{\partial \ln (\rho_{\rm g} c_{\rm s}^2)}{\partial \ln r}.
\end{eqnarray}
Here, $c_s$ is the isothermal sound velocity. The dimensionless stopping time, $St$, called the Stokes parameter, is 
given by  \citep{adachi76} 
\begin{equation}
 St = 
\left\{
\begin{array}{lcl}
\frac{3 m}{8 s^2 \Sigma_{\rm g}}& {\rm for}& \frac{s}{\lambda_{\rm mfp}} < \frac{9}{4},\\
%\frac{3 m \Omega}{8 \sqrt{2\pi} s^2 \rho_{\rm g} c_{\rm s} }& {\rm for}& s < \frac{9\lambda_{\rm mfp}}{4},\\
\frac{m}{6 s \lambda_{\rm mfp} \Sigma_{\rm g}} &{\rm for }&  \frac{9}{4} \leq \frac{s}{\lambda_{\rm mfp}} < \frac{12 h_{\rm g}}{\eta r},\\
% \frac{m \Omega }{6 \sqrt{2\pi} s \lambda_{\rm mfp} \rho_{\rm g} c_{\rm s} }&{\rm for }&  \frac{9\lambda_{\rm mfp}}{4} \leq s < \frac{12 \lambda_{\rm mfp} c_{\rm s}}{\eta r \Omega},\\
 \frac{4 m }{\pi s^2 \rho_{\rm g} \eta r }& {\rm for}& \frac{s}{\lambda_{\rm mfp}} \geq \frac{12 h_{\rm g}}{\eta r}, 
% \frac{4 m }{\pi s^2 \rho_{\rm g} \eta r }& {\rm for}& s \geq \frac{12 \lambda_{\rm mfp} c_{\rm s}}{\eta r \Omega}, 
\end{array}
\right.\label{eq:st}
\end{equation}
where $\rho_{\rm g}$ is the mid-plane gas density, $h_{\rm g} = c_{\rm s}/\Omega$ is the gas scale height, and $\lambda_{\rm mfp}$ is the mean free path.

As discussed above, collisional fragmentation is negligible for $St \la 1$. 
For further collisional growth, fragmentation is unimportant until planetary embryo formation \citep{kobayashi16,kobayashi18}. We ignore collisional fragmentation even after embryo formation because of the uncertainty of collisional outcome models. This crude assumption is good to compare with the studies for pebble accretion, in which collisional fragmentation is also ignored except for consideration of pebble sizes \citep{bitsch18,lambrecht19,johansen19}. In addition, collisional fragmentation for pebble formation works negatively for pebble accretion because of low $\varepsilon$ for $St \la 10^{-2}$ (Figure \ref{fig:epsi}). 
Therefore, we consider only the collisional merging. 

For $St \gtrsim 1$, the collisional kernel is scaled by the masses of bodies using the Hill radius ($r_{\rm H,1,2} = [(m_1+m_2)/3 M_*]^{1/3} r$): 
\begin{equation}
 K(m_1,m_2) = r_{\rm H,1,2}^2 {\cal P}(\tilde{e}_{1,2},\tilde{i}_{1,2}) \Omega, 
\label{eq:Kernel}
\end{equation}
where $\cal P$ is the collisional probability, and $\tilde{e}_{1,2}$ and $\tilde{i}_{1,2}$ are the Hill scaled relative eccentricity and inclination between $m_1$ and $m_2$. 
Sufficient mutual interaction between planetesimals results in the uniform orbital phases and eccentricities and inclinations following Rayleigh distributions \citep{ida_makino92}. Therefore, $\tilde{e}_{1,2} = (e_1^2 + e_2^2)^{1/2}r/r_{\rm H,1,2}$ and $\tilde{i}_{1,2} = (i_1^2 + i_2^2)^{1/2}r/r_{H,1,2}$, where $e_1$ and $i_1$ ($e_2$ and $i_2$) are the mean eccentricity and inclination of bodies with $m_1$ ($m_2$), respectively. The collisional probability $\cal{P}$ is given by the limiting solutions for $\tilde{e}_{1,2}, \tilde{i}_{1,2} \ll 1$, 
$\tilde{e}_{1,2},\tilde{i}_{1,2} \approx 0.2-2$, and 
$\tilde{e}_{1,2},\tilde{i}_{1,2} \gg 1$  \citep{inaba01}. 
In addition, 
we consider the enhancement of $\cal P$ due to planetary atmospheres 
 \citep{inaba_ikoma03}
and the strong gas drag around a massive planets \citep{ormel10b} 
The details of $\cal P$ are described in Appendixes \ref{app_enhance} and \ref{app_sc:p}. 

Collision-less interactions among bodies induce the evolution of their eccentricities and inclinations, which is sensitive to the mass spectrum of bodies. 
We calculate the $e$ and $i$ evolution together with the mass evolution, taking into account the mutual interaction between bodies such as viscous stirring and dynamical friction, gas drag, and the perturbation from the turbulent density fluctuation  \citep{kobayashi18}. 
The detailed treatment of $e$ and $i$ evolution is described in Appendix \ref{app_de_di}. 
We developed a simulation for planetesimal accretion ($St \gg 1$), which perfectly reproduces the result obtained from the direct $N$ body simulation  \citep{kobayashi+10}. 

%We expand the simulation for small $St$. 
%We describe the detailed treatment for $\cal P$ in Appendix. \ref{app_sc:p}. 
%Here we briefly explain the expanded treatment for $\cal P$. 
For $St \lesssim 1$, we calculate $\cal P$ 
additionally using the scale height and the relative velocity. 
For $St \ll 1$, the scale height for bodies with $m_1$ and $St_1$ is given by  \citep{youdin07}
\begin{equation}
 h_{\rm s,1} = h_{\rm g} 
\left(
1+ \frac{St_1}{\alpha_{\rm D} } \frac{1+ 2 St_1}{1+St_1}
\right)^{-1/2}, 
\end{equation}
where %$h_{\rm g} = c_{\rm s} / \Omega$ is the gas scale height, 
$\alpha_{\rm D}$ is the dimensionless turbulent parameter. 
We introduce the relative scale height between $m_1$ and $m_2$ as 
\begin{equation}
 h_{{\rm s},1,2} = [\pi ( h_{\rm s,1}^2 + h_{\rm s,1}^2)/2]^{1/2}. 
\end{equation}
The relative velocity is given by 
\begin{equation}
 v_{\rm rel,gas}^2 = \Delta v_{\rm B}^2 + \Delta v_r^2 + \Delta v_\theta^2 + \Delta v_z^2 + \Delta v_{\rm t}^2, 
\end{equation}
where $\Delta v_{\rm B}$, $\Delta v_r$, $\Delta v_\theta$, $\Delta v_z$, and $\Delta v_{\rm t}$ are the relative velocities induced by the Brownian motion, radial and azimuthal drifts, vertical settling, and turbulence, respectively 
(detailed description in Appendix \ref{sc:app_vrel}). 
%The drift and settling velocities are analytically given  \citep{adachi76}. We determine $\Delta v_{\rm t}$ using the analytic formulae for Kolmogorov turbulence analytically derived \citep{ormel07}. 
For $St_1 \ll 1$ and $St_2 \ll 1$, $K$ is expressed using $h_{{\rm s},1,2}$ and $v_{\rm rel,gas}$ as \citep{okuzumi12}, 
\begin{equation}
 K = \frac{\pi (s_1+s_2)^2 v_{\rm rel,gas}}{2 h_{\rm s,1,2}}.\label{eq:K_small} 
\end{equation}
We therefore expand Eq.~(\ref{eq:Kernel}) to apply the case for $St \lesssim 1$ 
using $h_{{\rm s},1,2}$ and $v_{\rm rel,gas}$. 

The collisional probability $\cal P$ is the function of $(\tilde{e}_{1,2}^2 + \tilde{i}_{1,2}^2)^{1/2}$ and $\tilde{i}_{1,2}$, which represent the relative velocity and the relative scale height, respectively. Therefore, 
we use the greater values of $(\tilde{e}_{1,2}^2 + \tilde{i}_{1,2}^2)^{1/2}$ or $v_{\rm rel,gas}/r_{\rm H,1,2} \Omega$ and $\tilde{i}_{1,2}$ or $h_{{\rm s},1,2}/r_{\rm H,1,2}$ for the function of $\cal P$. 
We then calculate the collisional Kernel $K$ for any $St$. Using this method, we calculate $K$ for $St_1, St_2 \ll 1$, which 
corresponds to Eq.~(\ref{eq:K_small}). 
Therefore, we apply this method for bodies from dust grains to planets.

\subsection{Bulk Density}
\label{sc:bulk_density}

The collisional growth of dust grains produces the fractal dust
aggregates, whose $\rho_{\rm b} = (3 m / 4 \pi
s^3)$ is lower than the original material. 
The stopping time $St$ depends on $\rho_{\rm b}$. 
The evolution of $\rho_{\rm b}$ is significantly important for collisional growth for $St \lesssim 1$. 
We model $\rho_{\rm b}$ as
\begin{equation}
 \rho_{\rm b} = \left[\rho_{\rm mat}^{-1} + (\rho_{\rm s} + \rho_{\rm m} + \rho_{\rm l})^{-1}\right]^{-1}, 
\end{equation}
where $\rho_{\rm mat} = 1.4 \,{\rm g/cm}^3$ is the material density, corresponding to the density of compact bodies or monomer grains in dust aggregates, 
\begin{eqnarray}
 \rho_{\rm s} &=& \rho_{\rm mat} \left(\frac{m}{m_{\rm mon}}\right)^{-0.58},\label{eq:rho_s}
\\
\rho_{\rm m} &=& 10^{-3} \,{\rm g/cm}^3,
\\
\rho_{\rm l} &=& \left(\frac{256 \pi G^3 \rho_{\rm mat}^9 s_{\rm mon}^9  m^{2}}{81 E_{\rm roll}^3 }\right)^{1/5},\label{eq:rho_l} 
\end{eqnarray}
$s_{\rm mon} = 0.1\,\micron$ is the monomer radius, $m_{\rm mon} = 4 \pi \rho_{\rm mat} s_{\rm mon}^3/3$ is the monomer mass, $E_{\rm roll}=4.74\times 10^{-9}\,{\rm erg}$ is the rolling energy between monomer grains, and $G$ is the gravitational constant. 

The densities $\rho_{\rm s}$, $\rho_{\rm m}$, and $\rho_{\rm l}$ almost correspond to the bulk density for small, intermediate, and large bodies, respectively. 
For small dust, collisional growth occurs without collisional compaction. Eq. (\ref{eq:rho_s}) is determined by the model given in the previous study  \citep{okuzumi12} under the assumption of the collisional evolution between same-mass bodies without collisional compaction, which is almost similar to the density evolution with the fractal dimension $\sim 2$  \citep{okuzumi12}. For large bodies, the bulk density increases with increasing mass by self-gravity compaction until compact bodies with $\rho_{\rm b} = \rho_{\rm mat}$ and the equilibrium density is given by Eq.~(\ref{eq:rho_l})  \citep{kataoka13}. 

\begin{figure}[htb]
\plotone{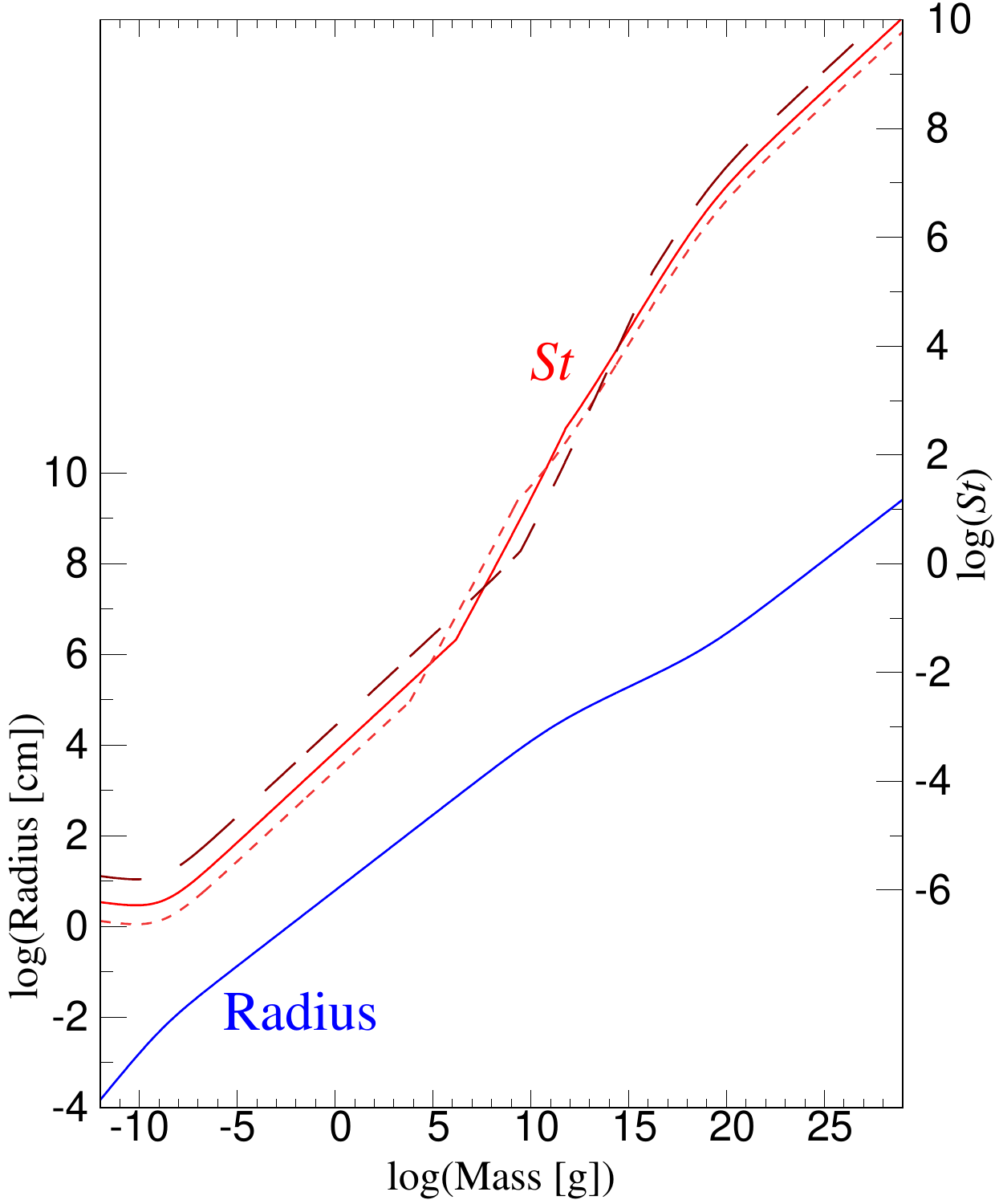}
%\plotone{mass_size.eps}
\caption{Radius or $St$ as a function of mass. The values of $St$ are given for $3$\,au (short dashed line), 6.8\,au (solid line), and 20\,au (long dashed line).
\label{fig:mass_size}
} 
\end{figure}

For intermediate bodies, the bulk density is most important for $St \sim 1$, which determined the fate of bodies. The bulk density is determined by the compression due to ram pressure of the disk gas  \citep{kataoka13}. 
We estimate $\rho_{\rm b}$ at $St = 1$ under the assumption of the Epstein gas drag, 
\begin{equation}
 \rho_{\rm b} \sim 6.3 \times 10^{-4} \left(\frac{r}{10\,{\rm au}}\right)^{-5/6}\,{\rm g/cm}^3.\label{eq:rho_gas} 
\end{equation}
It should be noted that 
$\rho_{\rm b}$ for $St = 1$ is smaller than that given by Eq.(\ref{eq:rho_gas}) at $r\lesssim 10$\,au because the Stokes gas drag is dominant for $St \sim 1$ at the inner disk. 
Therefore $\rho_{\rm b}$ for $St \sim 1$ becomes up to $\sim 10^{-3} {\rm g/cm}^3$ so that 
we simply choose the value of $\rho_{\rm m} = 10^{-3} \, {\rm g/cm}^3$ according to the estimate. 
Figure \ref{fig:mass_size} shows the radii or $St$ of bodies in the model as a function of mass. 

%\section*{Main}

\section{Result}
\label{sc:result}

We perform a DTPS for the collisional evolution of bodies drifting due to gas drag and Type I migration in a protoplanetary disk. 
We set a disk with the inner and outer radii of $\approx 3$\,au and $\approx 108$\,au, whose gas surface density is inversely proportional to $r$ (see Eq.~\ref{eq:sigmag}). 
%The dust to gas ratio of 0.017 is assumed according to the minimum-mass solar nebula model \citep{hayashi81}. 
The disk mass corresponds to $0.036 M_\sun$ (total solid mass $\approx 210 M_\oplus$), which is smaller than the required mass for the pebble accretion (see Eq.~\ref{m_disk_req}). 
%The initial total solid mass is about 210 Earth masses. 
Solid bodies initially have a mass $m = 5.9 \times 10^{-15}$\,g (corresponding to a radius of $0.1\,\micron$). 
%We use the bulk-density evolution model including the collisional and gravitational compaction effects \citep{okuzumi12,kataoka13}. 
We set the turbulent strength to be $\alpha_{\rm D} = 10^{-3}$.

\begin{figure*}[htb]
\plotone{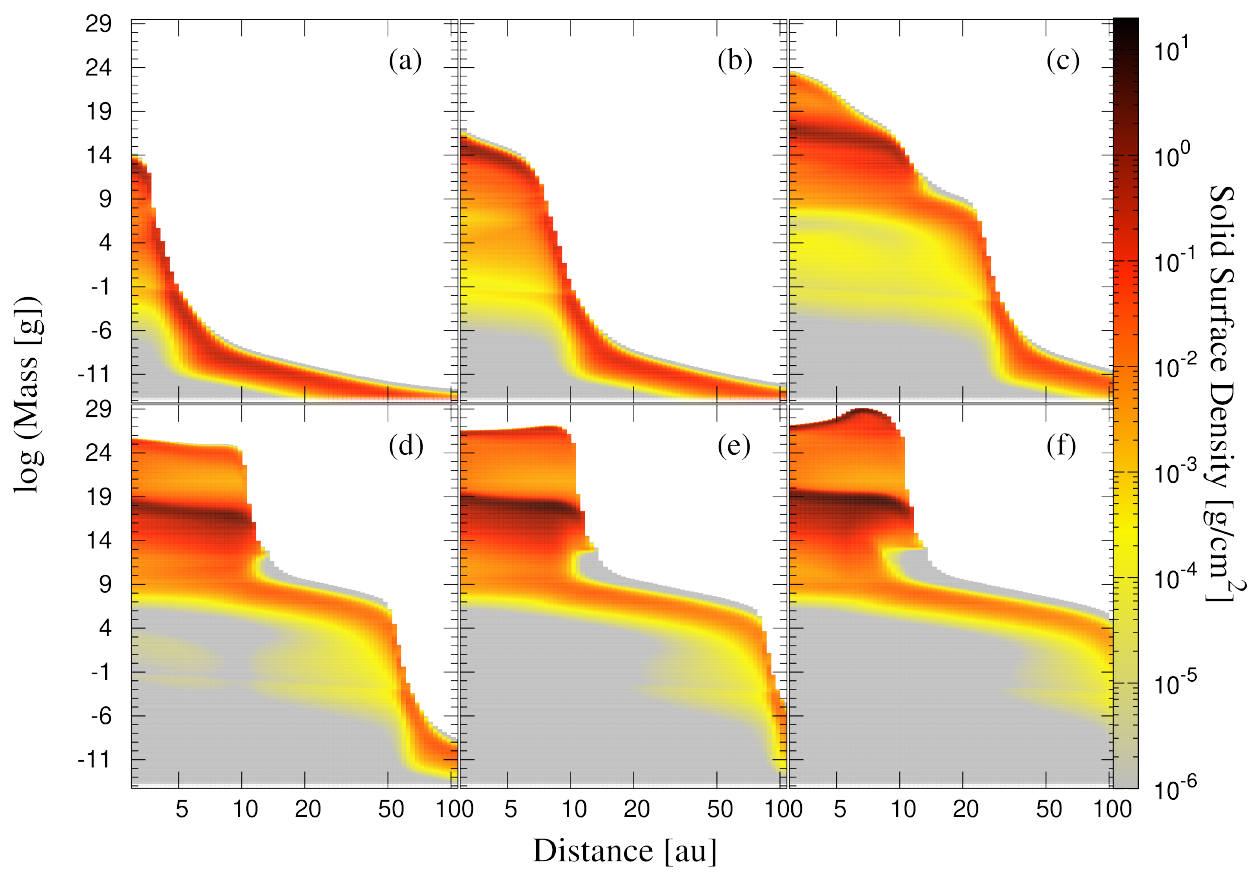}
%\plotone{multi.eps}
%\plotone{multi2.eps}
%\plotone{multi_test.pdf}
%\plotone{multi.pdf}
%\includegraphics[width={0.85\linewidth}]{test.pdf}% 
\caption{Solid surface density at $t=5.6 \times 10^2$ (a), $2.1 \times 10^3$ (b), $1.5 \times 10^4$ (c), $5.6 \times 10^4$ (d), $1.2 \times 10^5$ (e), $2.1 \times 10^5$ (f) years, as a function of the mass of bodies and the distance from the host star. The values of the solid surface density are shown in the color bar.
\label{fig:dist}
} 
\end{figure*}

Figure \ref{fig:dist} shows 
the surface density of bodies whose masses are similar to $m$ within about a factor of 2, as a function of $m$ and the distance from the host star. 
Dust growth occurs around $r \approx 5$\,au at $t \approx 560$ years (Figure \ref{fig:dist}a). Largest bodies reach at $m\sim 10^{13}$\,g at 3\,au. 
The drift of bodies is controlled by the gas coupling parameter of bodies $St$
(see Eq.~\ref{eq:st}). 
%, where $St$ is the stopping time multiplied by the Keplerian frequency \citep{adachi76}. 
Bodies have highest drift velocities at $m\sim 10^8$, corresponding to $St = 1$ (Figure \ref{fig:mass_size}). For low-density bodies, the collisional growth timescale is much shorter than the drift timescale so that large bodies with $St \gtrsim 1$ are formed via collision growth \citep{okuzumi12}. Dust collisional growth propagates from the inner to outer disk \citep[Figure \ref{fig:dist}b, see also][]{ohashi21}. Dust growth front reaches 20, 50, 90\,au and the outer boundary at $t \approx 1.5 \times 10^4$, $5.6 \times 10^4$, $1.2 \times 10^5$, and $2.1 \times 10^5$ years, respectively (Figure \ref{fig:dist}c--f). 
Radial drift is more dominant than collisional growth for bodies with $St \sim 1$ beyond 10\,au. The drifting bodies grow to planetesimals in the disk inside 10\,au. 

In the early growth (Figure \ref{fig:dist}a,b), the total solid surface densities are mainly determined by largest bodies. At $t \approx 6 \times 10^4$\,years (Figure \ref{fig:dist}c), the runaway growth of bodies with $m = 10^{13}$--$10^{16}$\,g occurs at $r\lesssim 6$\,au. The solid surface density of planetesimal-sized bodies ($m\sim 10^{18}$\,g) becomes dominant. Planetary embryos with $m\sim 10^{24}$\,g are formed at $r \lesssim 10\,$au via the runaway growth (Figure \ref{fig:dist}d). The further growth of embryos occurs via collisions with planetesimals (Figure \ref{fig:dist}e). The largest planetary embryos exceed 10 Earth masses even at $t \approx 2 \times 10^5$ years (Figure \ref{fig:dist}f). 

\begin{figure}[htb]
%\epsscale{0.5}
\plotone{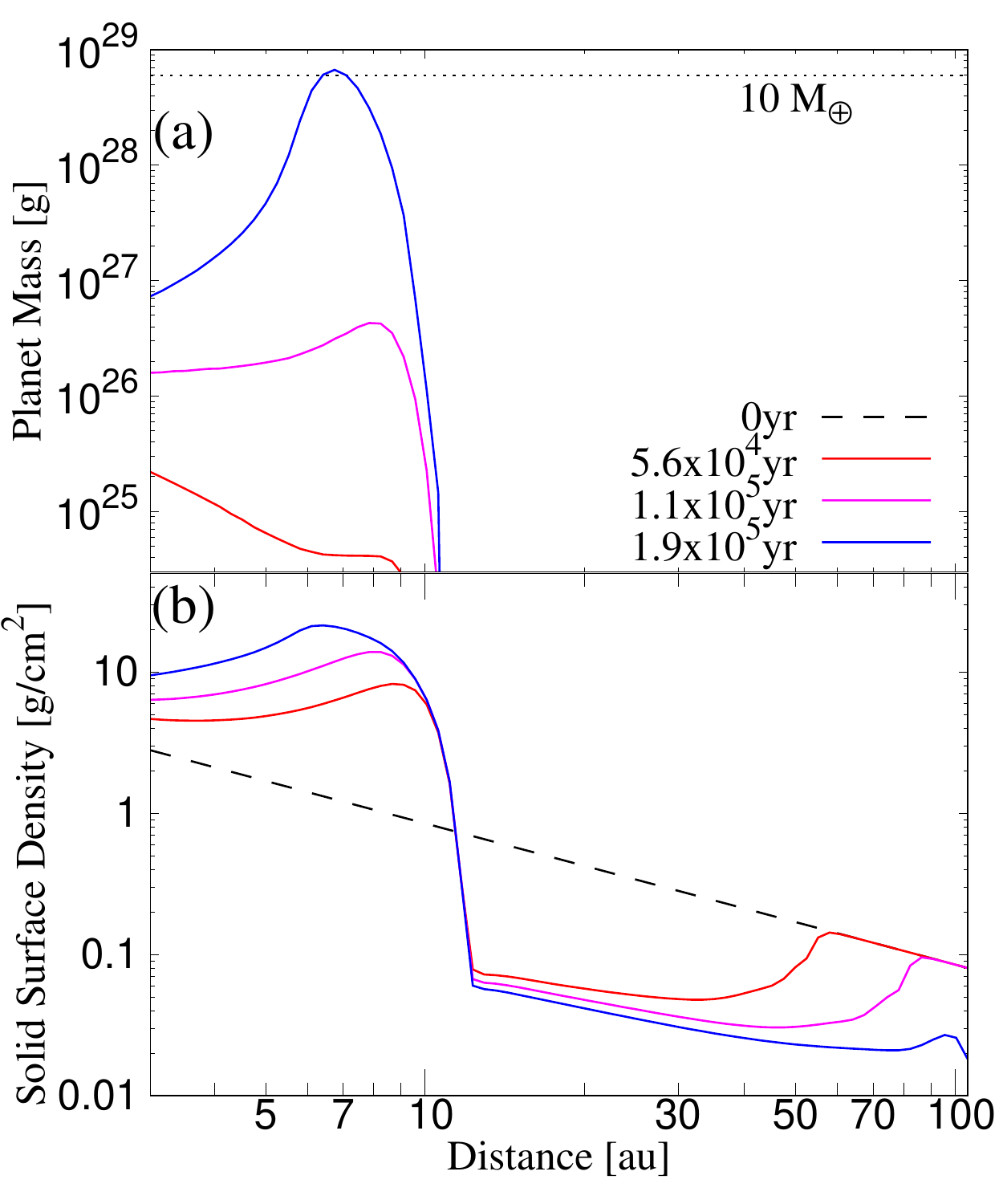}
%\plotone{fig_mp_sigma_port.eps}
%\includegraphics[width={0.425\linewidth}]{fig_mp_sigma_port.eps}% 
\caption{Planet mass (a) and surface density of bodies (b) as a function of the distance from the host star, where the planet mass is given by the mass of the largest bodies in each annulus in the disk. 
\label{fig:mp_sigma}
} 
\end{figure}

Collisional growth successfully forms bodies with $m \gg 10^{10}$g ($St \gg 1$) only at $r \la 10$\,au (Figures ~\ref{fig:dist}d--f). To overcome the drift barrier at $St \approx 1$, bodies with $St \approx 1$ should grow via collisions much faster than their radial drift. The requirement for this condition is that bodies with $St = 1$ feels gas drag in the Stokes regime \citep{okuzumi12}. Therefore, bodies for $St =1$ has $s \gg 9 \lambda_{\rm mfp}/4$ (see Eq.~\ref{eq:st}); 
\begin{equation}
 \frac{8 \Sigma_{\rm g}}{9 \pi \rho_{\rm b} \lambda_{\rm mfp}} \gg 1.\label{eq:stokes_condition} 
\end{equation}
For bulk densities and disk conditions given in \S~\ref{sc:bulk_density} and \ref{sc:mdisk}, Eq.~(\ref{eq:stokes_condition}) corresponds to $r_{\rm grow} \ll 24$\,au, where $r_{\rm grow}$ is the radius inside which pebbles can grow to planetesimals. Therefore, collisional growth results in planetesimals with $St \gg 1$ for $r \la 10$\,au. The radial drift of planetesimals with $St \gg 1$ is much slower than that of pebbles $St \la 1$; the pile-up results in the enhancement of solid surface densities at $r \la 10$\,au (Figures \ref{fig:dist}d--f and \ref{fig:mp_sigma}b). Pebbles formed in the whole disk with the total solid mass $M_{\rm solid, disk}$ finally drift inward across $r_{\rm grow}$, so that the enhanced surface density is estimated to be $M_{\rm solid,disk}/\pi r_{\rm grow}^2 \approx 18 (r_{\rm grow}/10\,{\rm au})^{-2} \,{\rm g \, cm}^{-2}$ (compare with Figure \ref{fig:mp_sigma}b). 

Figure~\ref{fig:mp_sigma}a shows the mass of largest planetary embryos in each annulus of the disk. Planetary embryos acquire $10 \, M_\oplus$ around $r \approx 6$--7\,au at $t \approx 2 \times 10^5$ years. Such rapid formation of massive embryos is achieved via the pile-up of bodies in $r < 10$\,au (Figure \ref{fig:mp_sigma}b). As explained above, bodies with $St < 1$ drift inwards from the outer disks until the bodies grow to $St \gg 1$ in $r < 10$\,au. 
%The enhancement of $\Sigma_{\rm s}$ increases 
%the isolation mass, $M_{\rm iso}$, with which cores fully accrete the surrounding bodies, estimated to be about 100 Earth masses at 7\,au for 
%$\Sigma_{\rm s} = 15 \,{\rm g \, cm}^{-2}$. 
The solid surface density increases to $20 \,{\rm g \, cm}^{-2}$ at 7\,au in $2\times 10^5$ years (see Figure \ref{fig:mp_sigma}b), the formation of cores with $10\,M_\oplus$ requires the surface density of $3 \,{\rm g \, cm}^{-2}$. 
Therefore, only about 15\,\% of bodies are needed for the core formation in the enhanced disk. 
The $\Sigma_{\rm s}$ enhancement effectively accelerates the growth of cores. However, the 
growth rate depends on the mass spectrum of bodies accreting onto cores. 

\begin{figure}[htb]
%\epsscale{0.5}
\plotone{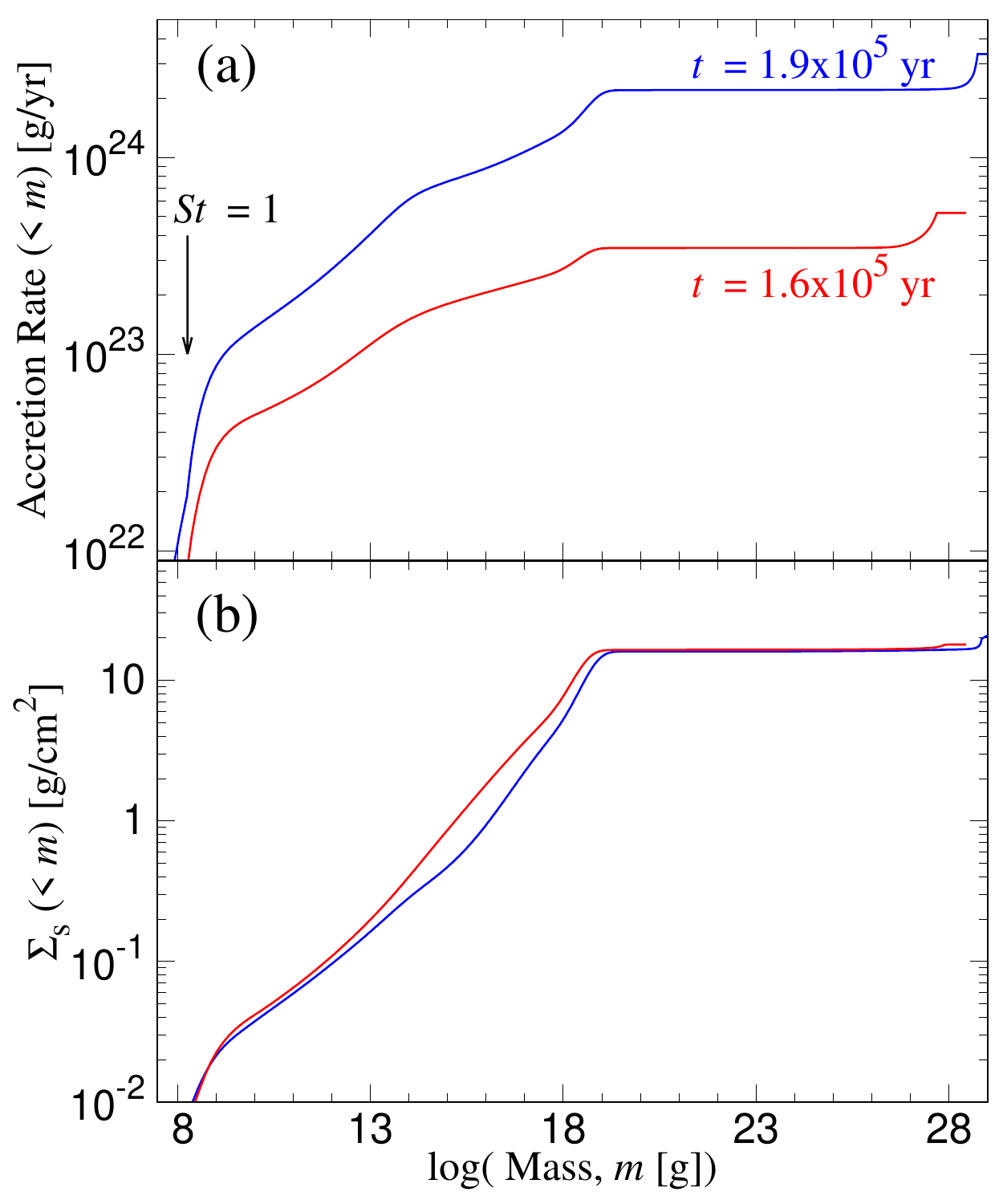}
%\plotone{accretion_rate_6_75au_port.eps}
%\includegraphics[width={0.425\linewidth}]{accretion_rate_6.75au_port.eps}% 
\figcaption{(a) Cumulative accretion rate of bodies, whose masses are smaller than $m$, onto a largest core in the annulus at 6.75\,au as a function of mass $m$. (b) Cumulative solid surface density of bodies with masses smaller than $m$ at 6.75\,au. 
\label{fig:accretion_rate}
} 
\end{figure}

We additionally investigate which masses of bodies mostly contribute to the core growth. 
The cumulative accretion rate of bodies onto the largest core in the annulus at $r= 6.75$\,au is shown in Figure \ref{fig:accretion_rate}a.
The contribution of pebbles ($St < 1$) to the accretion rate is minor, because the solid surface density of bodies with $St < 1$ is tiny (Figure \ref{fig:accretion_rate}b). Collisions with 100\,m--10\,km sized bodies of $m=10^{9}$--$10^{19}$\,g mainly contribute to the accretion rate, while the solid surface density is mainly determined by planetesimal-sized bodies of $m \sim 10^{19}$\,g (Figure \ref{fig:accretion_rate} and see also Figure ~\ref{fig:dist}). The atmospheric collisional enhancement promotes the accretion of sub-kilometer-sized bodies of $m\sim 10^{16}$\,g, which are in the course of growing to planetesimals. The growth of planetary cores additionally increases their Hill radii so that collisions between planetary embryos occur. The embryo accretion therefore increases the total accretion rate by a factor of about 1.5 additionally \citep{chambers06,kobayashi+10}. 

%A planetary core with mass $M$ and radius $R_{\rm p}$ grows via the collisional accretion of planetesimals of mass $m$ with the surface density $\Sigma_{\rm s}$. 
%\begin{equation}
% t_{\rm grow} \approx \frac{v_{\rm rel}^2}{2 \pi G R_{\rm e} \Sigma_{\rm s} \Omega}, 
%\end{equation}
%where $v_{\rm rel}$ is the relative velocity between the core and planetesimals, $R_{\rm e}$ is the effective collisional radius of the core, and $\Omega$ is the orbital frequency. 
%The relative velocity is given by the equilibrium between the planetary stirring and the gas drag and the effective radius $R_{\rm e}$ is enhanced by a factor 3 due to the core atmosphere for $m=10^{19}$\,{g}  \citep{kobayashi11}. If we use $\Sigma_{\rm s} = 1.2\,{\rm g \,cm}^{-2}$ of the initial condition, we estimate $t_{\rm grow} \sim 10^7$\,years. However, 

We again estimate the growth timescale of cores via planetesimal accretion 
in this condition using Eq.~(\ref{eq:tg}) in \S~\ref{sc:timescale}. 
The solid surface density of planetesimals or planetesimal precursors increases to $15\,{\rm g/cm}^2$ (Figure \ref{fig:accretion_rate}b). 
As mentioned above, the contribution of planetesimal precursors to the accretion is significant. The enhancement factor $R_{\rm e}/R_{\rm p}$ proportional to $m^{-1/9}$ is higher for small planetesimals \citep{kobayashi11}. The growth timescale is then estimated to be $t_{\rm grow} \approx 1.5 \times 10^4 (\Sigma_{\rm s} /15\,{\rm g\,cm}^{-2})(m/10^{16}\,{\rm g})^{11/25}$ years, corresponding to the accretion rate of $4.0 \times 10^{24}$\,g/yr for $M_{\rm p} = 10 M_\oplus$. This value is consistent with the accretion rate at $t \approx 1.9 \times 10^5$ years (see Figure \ref{fig:accretion_rate}a). 

The accretion of planetesimals with mass $m = 10^{15}$--$10^{19}$\,g induces the rapid growth of planetary embryos. Such planetesimals are produced via collisional growth of planetesimal precursors with $m \sim 10^8 - 10^{15}$\,g. 
The bulk density of such planetesimal precursors for $m \gtrsim 10^{13}$\,g 
is given by $\rho_{\rm b} \propto m^{2/5}$ (see \S~\ref{sc:bulk_density}). 
Their collisional timescale among planetesimal precursors with mass $m$ and radius $r_{\rm p}$ is given by 
\begin{equation}
 t_{\rm col} \approx \frac{m}{\pi r_{\rm p}^2 \Sigma_{\rm s} \Omega}, 
\label{eq:tcol} 
\end{equation}
which is estimated to be $t_{\rm col} \approx 4.5 \times 10^3$ years at 7\,au for $m=10^{13}\,$g and $\Sigma_{\rm s} = 0.2 \,{\rm g \, cm}^{-2}$ according to Figure \ref{fig:accretion_rate}b. 
On the other hand, the drift timescale of planetesimal precursors is given by 
\begin{equation}
 t_{\rm drift} \approx \frac{St}{2 \eta \Omega},  
\label{eq:tdrift}
\end{equation}
where $\eta$ is the dimensionless parameter depending on the pressure gradient. 
The drift timescale is estimated to be $t_{\rm drift} \approx 3.5 \times 10^5$ years for $m=10^{13}$\,g. Therefore, $t_{\rm col} / t_{\rm drift} \approx 1.3 \times 10^{-2} (\Sigma_{\rm s} /0.2 \,{\rm g \, cm}^{-2})^{-1} \ll 1$ independent of $m$. 
For such planetesimal precursors, the collisional growth timescale is much shorter than the drift timescale.

Pebbles with $St \sim 0.1$ drift from the outer disk, and the collisional growth among the pebbles produces planetesimal precursors prior to their drift. 
Because $t_{\rm col} \ll t_{\rm drift}$ as estimated above, 
planetesimal precursors grow without significant drift in the inner disk ($<10\,$au) until gravitational scatterings by planetary cores, which 
lead to the uniform distribution of planetesimal precursors around cores. 
%$t_{\rm col}$ is comparable to the system time, 
%resulting in the formation of planetesimals. 
The surface density of planetesimal precursors is much smaller than that of planetesimals (Figure \ref{fig:accretion_rate}b). Planetesimal precursors are maintained via the supply from the growth of pebbles drifting from the outer disk. This mechanism leads to the sustainable accretion of small planetesimals, 
resulting in the rapid growth within 0.2\,Myr. 

If the scattering of planetesimal precursors by a solid core is
comparable to the drift, such solid cores may open gaps up in a
planetesimal-precursor disk, which would reduce the accretion rate of small
planetesimals  \citep{levison10}. However, the collisional growth timescale of planetesimal precursors is much shorter than the gap opening timescale comparable to $t_{\rm drift}$, 
so that planetesimals are then formed via collisonal growth of precursors prior to gap opening.  Therefore,
the rapid growth is achieved without the gap opening in the solid disk.

\section{Discussion}
\label{sc:discussion}

We show the rapid core formation at 6--7\,au in $2\times 10^5$\,years.  
We obtain the similar result for a weak turbulent level of $\alpha_{\rm D} = 10^{-4}$, with which the simulation results in the core formation at $\approx 7$\,au in $3 \times 10^5$ years. 
A sufficient massive 
core starts gas accretion and type II migration. 
The orbital samimajor axis of a gas
giant with Jupiter mass resulting from the gas accretion and type II
migration is about 0.9 times that of the original core  \citep{tanaka20}. 
Therefore, the
first gas giant is formed around 6\,au. The giant planets in the solar
system may experience migration.  Outward migration of Neptune can
explain the orbital eccentricities of Plutinos in the Kuiper belt. The exchange of angular momentum between Nepture and Jupiter
via interactions with planetesimals requires the original orbit of
Jupiter at $\sim 6$\,au  \citep{minton}. Therefore, the
formation of a core at 6--7\,au is consistent with the origin of
Jupiter.

On the other hand, the formation location of gas-giant cores depends on the solid/gas ratio, which is set to 0.017 in the simulation according to the minimum-mass solar nebula model  \citep{hayashi81}. We additionally carry out a simulation for the solid/gas ratio of 0.01, whose disk includes solids $\approx 120\,M_\oplus$.
The core formation occurs at $3\times 10^5$\,years at 3--4\,au. The population of exoplanets found via radial velocity surveys is high around 2--3 \,au  \citep{fernandes19}. The occurrence location of gas giants may be explained by the typical solid/gas ratio of $\sim 0.01$. 

For much smaller solid/gas ratios (solid masses smaller than $100 M_\oplus$), the growth timescales of cores are longer than the migration timescales, resulting in the difficulty of gas giant formation. In addition, small disk masses tend to make the planetesimal forming radii $r_{\rm grow}$ small. If $r_{\rm grow}$ is smaller than the snow line, the enhancement of solid surface density due to the mechanisms shown in \S.~\ref{sc:result} is less effective, depending on the outcomes of the sublimation of icy pebbles. Furthermore, disk sizes are also important. If disk sizes are much smaller than 100\,AU, pebble supply is stalled prior to core formation. Such difficulties of core formation may be important to discuss the gas giant occurrence rate \citep[$\sim 10 \%$,][]{mayor11}.

\section{Summary}
\label{sc:summary}

Gas giant planets are formed via rapid gas accretion of massive solid cores prior to type I planetary migration of cores (timescale of $\sim 10^5$ years). 
Core formation via accretion of 10 km sized or larger planetesimals requires times much longer than the migration timescale. Pebbles form via collisional coagulation in outer disks, which drift into the inner disk. The core-growth timescale via pebble accretion is much shorter than that via the accretion of 10 km sized planetesimals. However, pebble accretion mostly losses drifting pebbles, which requires more than $300\,M_\oplus$ in solid for single-core formation (\S. \ref{sc:estimate}). 
To reconcile the issue, we need to consider collisional growth of pebbles to planetesimals. 

We thus develop a simulation for the collisional evolution of bodies 
from dust to planet (“Dust-to-planet” simulation; Hereafter, DTPS), which 
describes both planetary growth modes via planetesimal 
accumulation and pebble accretion (\S~\ref{sc:DTPSmodel}). 
The result of the DTPS shows as follows. 

\begin{itemize}
 \item[1.] Collisional coagulation of dust aggregates forms planetesimals 
in the inner disk ($\la 10\,$au) thanks to the realistic bulk densities of dust
aggregates, while pebbles drift inwards from the outer disk ($\ga 10\,$au).
Planetesimal formation via collisional growth of pebbles 
increases the solid surface density at 6--9 \,au by the factor of $\sim 10$ (Figure \ref{fig:mp_sigma}).

\item[2.]
Planetesimals growing from pebbles are relatively small, whose accretion
rate to planetary cores per unit surface density is relatively
high. Both the enhanced solid surface density and the high accretion
rate accelerate the growth of cores significantly. The accretion
rate becomes $4 \times 10^{24}$\,g at about $7$\,au in $2\times 10^5$ years (Figure \ref{fig:accretion_rate}), corresponding to the core-growth timescale of $2 \times 10^4$\,years, which is much shorter than Type I migration times of cores.  Solid cores
are formed in 6--7\,au without significant type I migration, which are
likely for the formation of Jupiter in the Solar System.

\item[3.]
Core formation with only pebble accretion requires very massive solids in disks ($> 300 M_\oplus$, as estimated in S.~\ref{sc:mdisk}), which are very rare among observed protoplanetary disks.
Our dust-to-planet simulations (DTPS) show 
a disk with the total solid mass of $210 M_\oplus$ produces a gas giant orbiting at $\approx 5$\,AU, while inner gas giant planets are formed in less massive disks each with a solid mass of $> 100 M_\oplus$. 
Thus our model naturally explains the formation of gas giant planets from protoplanetary disks each with a solid mass of $\approx 100$--200\,$M_\oplus$. 
\end{itemize}

\begin{acknowledgements}
We thank Hiroshi Kimura for useful conversation about dust properties. 
HK is grateful to Hana Nishikawa for helpful discussion at the begging of this project. 
The work is supported by
Grants-in-Aid for Scientific Research
(17K05632, 17H01105, 17H01103, 18H05436, 18H05438, 20H04612, 21K03642)
from MEXT of Japan. 
\end{acknowledgements}

\appendix

\section{radial drift}
\label{sc:app_drift}

For $St \gg 1$, the drift velocity induced by gas drag is given by 
\citep{adachi76,inaba01}
\begin{equation}
 v_{\rm drag} = - \frac{2 r \Omega}{St} \left\{
\left[
\frac{E(3/4)+K(3/4)}{3\pi}
\right]^2
e^2 + 
\frac{4 i^2 
}{\pi^2}
+ \eta^2
\right\}^{1/2}, 
\end{equation}
where $K$ and $E$ are the complete elliptic integrals of the first and second kinds, and 
we ignore the higher order terms of $e$ and $i$ because the terms are small enough \citep{kobayashi15}. 
On the other hand, for $St \la 1$ 
$v_{\rm drag}$ is given by \citep{adachi76}
\begin{equation}
 v_{\rm drag} = - \frac{2 \eta r \Omega St}{1+St^2}. 
\end{equation}
We combine the both regimes as \citep{kobayashi+10}, 
\begin{equation}
 v_{\rm drag} = - \frac{2 r \Omega St}{1+St^2}
\left\{
\left[
\frac{E(3/4)+K(3/4)}{3\pi}
\right]^2
e^2 + 
\frac{4 i^2 
}{\pi^2}
+ \eta^2
\right\}^{1/2}. 
\end{equation}

\section{Atmospheric Enhancement}
\label{app_enhance}

The collisional radius is enhanced by an atmosphere. 
\citet{inaba_ikoma03} derived the enhancement factor for the radius, 
\begin{equation}
 \xi = \frac{3}{2} \frac{v_{\rm rel}^2 + 2 G m / s}{v_{\rm rel}^2 + 2 G m/r_{\rm H}} \frac{\rho_{\rm a}}{\rho_{\rm b}}, 
\end{equation}
where $v_{\rm rel}$ is the relative velocity, $r_{\rm H} = (m/3M_*)^{1/3}r$ is the Hill radius, $\rho_{\rm a}$ is the atmospheric density at $\xi s$ from the center of the body. 

We derive $\rho_{\rm a}$ according to \citet{inaba_ikoma03}. 
The pressure $P_{\rm a}$, temperature $T_{\rm a}$, and density $\rho_{\rm a}$ of the atmosphere at the distance $r_{\rm a}$ from the body have the relations as follows. 
\begin{eqnarray}
 \tilde{P}_{a} &=& \tilde{\rho}_{\rm a} \tilde{T}_{\rm a}, \\
 \tilde{T}_{\rm a}^4 &=& 1 + W_0 (\tilde P_{\rm a} -1),\\
 \tilde{r}_{\rm a} &=& 1 + \frac{1}{V_0} [4 (\tilde T_{\rm a} -1) + f(\tilde T_{\rm a},w_0)], 
\end{eqnarray}
where the dimensionless pressure $\tilde P_{\rm a} = P_{\rm a}/P_{\rm o}$, temperature $\tilde T_{\rm a} = T_{\rm a}/T_{\rm o}$, density $\tilde \rho_{\rm a} = \rho_{\rm a}/\rho_{\rm o}$, distance $\tilde r_{\rm a} = r_{\rm a}/r_{\rm o}$ are scaled by the pressure, temperature, density, and distance at the outer boundary, respectively, 
\begin{eqnarray}
 V_0 &=& \frac{Gm\rho_{\rm o}}{r_{\rm o} P_{\rm o}},\\
 W_0 &=& \frac{3 \kappa_{\rm a} L_{\rm a} P_{\rm o}}{4 \pi a_{\rm r} c G m T_{\rm o}^4},\\
 w_0 &=& |1-W_0|^{1/4}, \\
 g(\tilde T_{\rm a},w_0) &=& 
  \left\{
   \begin{array}{l l l}
    w_0 \ln\left(\frac{\tilde T_{\rm a}-w_0}{\tilde T_{\rm a}+w_0} \frac{1+w_0}{1-w_0}\right)	    	   
     \left(\arctan \frac{\tilde T_{\rm a}}{w_0} - \arctan \frac{1}{w_0} \right)
& \quad {\rm for} & W_0 < 1,
\\
\frac{w_0}{\sqrt{2}} 
 \left[
  \ln \left( \frac{\tilde T_{\rm a}^2 + \sqrt{2} w_0 \tilde T_{\rm a}+ w_0^2}
       {\tilde T_{\rm a}^2 - \sqrt{2} w_0 \tilde T_{\rm a} + w_0^2} 
       \frac{ 1 - \sqrt{2} w_0 + w_0^2}{1 + \sqrt{2} w_0 + w_0^2}        
       \right)
  + 2\left(
      \arctan \frac{\sqrt{2} w_0 \tilde{T}_{\rm a}}{w_0^2 - \tilde T_{\rm a}^2}
      - \arctan \frac{\sqrt{2} w_0 }{w_0^2 - 1}
     \right)
 \right]
& \quad {\rm for} & W_0 \geq 1,
   \end{array}
  \right.
\nonumber
\\&&
\end{eqnarray}
$a_{\rm r}$ is the radiation density constant, 
$\kappa_{\rm a}$ is the opacity, and $L_{\rm a}$ is the luminosity. 

The luminosity is given by 
\begin{equation}
 L_{\rm a} = {\rm MAX} \left(
\frac{G m \dot m}{s},4 \pi s^2 \sigma_{\rm SB} T^4 \right), 
\end{equation}
where $\dot m$ is the accretion rate of the body, 
$\sigma_{\rm SB}$ is the Stefan-Boltzmann constant, 
and the function ${\rm MAX}(x,y)$ gives the larger of $x$ and $y$. 

We set the outer boundary values and opacity as follows. 
\begin{eqnarray}
T_{\rm o} = T, P_{\rm o} = P, r_{\rm o} = {\rm MIN}(\frac{G m}{k_{\rm hc} c_{\rm s}^2},r_{\rm H}), \kappa_{\rm a} = 4 \zeta +0.01 {\rm cm}^2 {\rm g}^{-1} &\quad {\rm for}& \quad T < T_{\rm a} \leq 170 {\rm K},\\
 T_{\rm o} = 170\,{\rm K}, P_{\rm o} = P_{{\rm a},170\,{\rm K}}, r_{\rm o} = r_{\rm a,170\,K}, \kappa_{\rm a} = 2 \zeta +0.01 {\rm cm}^2 {\rm g}^{-1} &\quad {\rm for}& \quad 170\,{\rm K} < T_{\rm a} \leq 1700 {\rm K},\\
 T_{\rm o} = 1700\,{\rm K}, P_{\rm o} = P_{{\rm a},1700\,{\rm K}}, r_{\rm o} = r_{\rm a,1700\,K}, \kappa_{\rm a} = 0.01 {\rm cm}^2 {\rm g}^{-1} &\quad {\rm for}& \quad  T_{\rm a} > 1700 {\rm K},
\end{eqnarray}
where $P$ is the gas pressure at the disk mid-plane, $c_{\rm s}$ is the isothermal sound velocity, $k_{\rm hc}$ is the heat capacity ratio, 
the function ${\rm MIN}(x,y)$ gives the smaller of $x$ and $y$, the subscripts of $170\,{\rm K}$ and $1700\,{\rm K}$ indicate the values at $T_{\rm a} = 170$\,K and 1700\,K, respectively, and $\zeta$ is the reduction factor of the atmospheric opacity. A massive planetary body acquire an atmosphere. Small dust grains decrease until the formation of massive bodies. We therefore apply $\zeta = 10^{-4}$. 

\section{Collisional Probability}
\label{app_sc:p}

Taking into account the relative velocity induced by gas, we use 
\begin{equation}
 \tilde{e}^* = {\rm MAX} (v_{\rm rel,gas}/h_{1,2} r\Omega,\tilde{e}_{1,2}).
\end{equation}
We then introduce 
\begin{equation}
 \begin{array}{lcl}
  I &=& \left\{
   \begin{array}{lcr}
    \tilde{i}_{1,2} / \tilde{e}^* & {\rm for}&  v_{\rm rel,gas}/h_{1,2} r\Omega \leq \tilde{e}_{1,2},\\
    0.812 %0.811509 
& {\rm for}&  v_{\rm rel,gas}/h_{1,2} r\Omega > \tilde{e}_{1,2},\\
   \end{array}
	\right.
  \\
  I_{\rm t} &=& {\rm MAX}(h_{{\rm s},1,2}/rh_{1,2}, \tilde{i}_{1,2}) / \tilde{e}^*, 
 \end{array}
\end{equation}
where $h_{\rm 1,2} = r_{\rm H,1,2}$ is the dimensionless mutual Hill radius. 

The formulae of collisional probabilities for $St_1, St_2 \gg 1$ 
are mainly modeled by \citet{inaba01} \citep[see also][]{ormel_kobayashi12}. 
Using $I$ and $I_{\rm t}$, we modify the formula for $m_1 > m_2$ as 
\begin{equation}
 {\cal P} = \left\{ 
\begin{array}{l l}
{\rm MIN}({\cal P}_{\rm mid},
({\cal P}_{\rm low}^{-2} + {\cal P}_{\rm
high}^{-2})^{-1/2}) \quad {\rm for} & {\rm MIN}({\cal P}_{\rm high},{\cal P}_{\rm high}) > {\cal P}_{\rm low}, \\
{\cal P}_{\rm vl} & {\rm otherwise}, 
\end{array}
\label{eq:app_p} 
\right.
\end{equation}
where 
\begin{eqnarray}
 {\cal P}_{\rm low} &=& 11.3 \sqrt{{\tilde s}_{1,2}}, \\
 {\cal P}_{\rm mid} &=& \frac{{\tilde s}_{1,2}}{4 \pi {\tilde I}_{\rm t} \tilde{e}^*}
  \left(17.3 + \frac{232}{{\tilde s}_{1,2}}\right), \\
  {\cal P}_{\rm high} &=& \frac{{\tilde s}_{1,2}^2}{2\pi}
   \left(
    F(I,I_{\rm t}) + \frac{6 G(I,I_{\rm t})}{{\tilde s}_{1,2} {\tilde e}^{*2}}
       \right), \label{eq:app_p_high}
\\
 {\cal P}_{\rm vl} &=& \left\{
  \begin{array}{lcl}
   2 b_{\rm set}  
    \left(\frac{3}{2} b_{\rm set}  + \frac{\eta}{h_{1,2}}\right)
&{\rm for}& St_2 < {\rm MIN}(1,12 h_{1,2}^3/\eta^3), 
\\
   {\cal P}_{\rm low} + \frac{6.4}{St_2} &{\rm for}& St_2 > {\rm
    MAX}\left(\frac{\eta}{h_{1,2}},1\right),    
   \\
   2 {\rm MAX}(b_{\rm hyp},b_{\rm set}) v_{\rm a} &&{\rm otherwise.}
  \end{array}
\right. 
\end{eqnarray}
The dimensionless colliding radius of bodies 
$\tilde s_{1,2}$ is given by $(\xi_1 s_1+s_2)/h_{1,2}$ with the enhancement factor $\xi_1$ due to planetary atmospheres given in Appendix \ref{app_enhance}. 
The formula $\cal P_{\rm vl}$ is obtained by \citet{ormel10b}, where
$b_{\rm set}$, $b_{\rm hyp}$, and $v_{\rm a}$ 
are given by the solution of $b_{\rm set}^2 (b_{\rm set}+2 \eta / 3
h_{1,2}) = St_2$ with the factor of $\exp[ -(St_2 h_{1,2}^3/ 12
\eta^3)^{0.65}]$, ${\tilde s}_{1,2} \sqrt{1+6/{\tilde s}_{1,2} v_{\rm
a}^2}$, and $\eta \sqrt{1+4 St_2^2} / h_{1,2} (1+St_2^2)$,
respectively. 

The functions $F$ and $G$ are originally formulated by \citet{greenzweig92}. Taking into account $I_{\rm t}$, we modify them as 
\begin{eqnarray}
 F(I,I_{\rm t}) &=& \frac{\sqrt{3} \pi E(\overline{k}_F)}{I_{\rm t} \overline{k}_F} 
  \left\{
   1+ \frac{\overline{k^2}_F - \overline{k}_F^2}{2 \overline{k}_F^2} 
   \left(
    \frac{K(\overline{k}_F)}{E(\overline{k}_F)} - \frac{1}{1- \overline{k}_F^2}
   \right)
  \right\},\label{eq:F} \\
 G(I,I_{\rm t}) &=& \frac{2 \sqrt{3} \pi K(\overline{k}_G)}{I_{\rm t} (1+I)\overline{k}_G} 
  \left\{
   1+ \frac{\overline{k^2}_G - \overline{k}_G^2}{2 \overline{k}_G^2 (1-\overline{k}_G^2 )} 
   \left(
    1 - 2 \overline{k}_G^2
    - \frac{E(\overline{k}_G)}{K(\overline{k}_G)} 
    \frac{1 - 3 \overline{k}_G^2}{1- \overline{k}_G^2}
   \right)
  \right\},\label{eq:G} 
\end{eqnarray}
where $K$ and $E$ are the complete elliptic integrals of the first and second kinds, respectively, $\overline{k}_F = \sqrt{3} \pi / 4 I (1 + \arctan \sqrt{I^{-2}-1} / I \sqrt{1-I^2})$ for $I < 1$, $\sqrt{3} \pi /8 $ for $I=1$, and $\sqrt{3} \pi / 4 I (1 + \ln \sqrt{I^{2}-1} / I \sqrt{I^2-1})$ for $I > 1$, and 
$\overline{k^2}_F = 3 (1- 2 I^2 (1- 4 I \overline{k}_F / \sqrt{3} \pi)) / 4 (1-I^2)$ for $I \neq 1$ and $1/2$ for $I = 1$, 
$\overline{k}_{G} = \sqrt{3} \pi \sqrt{1-I}/ 4 \sqrt{1+I} \arctan (I^{-2}-1)$ for $I < 1$, $\sqrt{3} \pi/8$ for $I=1$, and $\sqrt{3} \pi \sqrt{I-1}/ 4 \sqrt{1+I} \ln (I^{2}-1)$ for $I > 1$, and $\overline{k^2}_G = 3 (1 - 4 \overline{k}_G (1+I)I / \sqrt{3} \pi) / 4 (1-I^2)$ for $I \neq 1$ and $1/2$ for $I=1$.
Our modifications for $F$ and $G$ are only $I_{\rm t}$ in the denominators of the first terms in Eqs.(\ref{eq:F}) and (\ref{eq:G}). 

If $I_{\rm t} = I$, the formulae of Eqs.(\ref{eq:app_p})--(\ref{eq:app_p_high}) are the same as those in \citep{inaba01}. In the limit of $St_1, St_2 \ll 1$, 
$\tilde{e}^* = v_{\rm rel,gas}/ r h_{1,2} \Omega \gg 1$ and $I_{\rm t} = h_{{\rm s},1,2} \Omega / r v_{\rm rel,gas}$, and $I = 0.812$. Therefore, $\cal P$ reduces to 
\begin{equation}
 {\cal P} \approx 1.57 \tilde{s}_{1,2}^2 v_{\rm rel}/ h_{{\rm s},1,2}. 
\end{equation}
This collisional probability corresponds to that between dust grains
with $St_1, St_2 \ll 1$.

\section{random velocity evolution}
\label{app_de_di}

For $St \gg 1$, collisional evolution depends on $e$ and $i$. 
We consider the $e$ and $i$ evolution due to gravitational interaction \citep{ohtsuki02}, gas drag \citep{adachi76}, and collisional damping \citep{ohtsuki92}. 
On the other hand, a body have a orbit determined by the Kepler law for $St \la 1$. The orbital elements of $e$ and $i$ do not indicate the motion of bodies. However, the collisional velocity is determined by $v_{\rm rel,gas}$ instead of $e$ and $i$. Therefore, we calculate the $e$ and $i$ evolution via the following equations. 
\begin{eqnarray}
 \frac{d e^2}{d t} &=& 
\left\{
\begin{array}{l l l}
 0 & \quad {\rm for} & \quad St <1, \\
\left.\frac{d e^2}{d t}\right|_{\rm i}+\left.\frac{d e^2}{d t}\right|_{\rm g}+
\left.\frac{d e^2}{d t}\right|_{\rm c}
 & \quad {\rm for} & \quad St \ge 1, 
\end{array}
\right.
\label{de2dt}
\\
  \frac{d i^2}{d t} &=& 
\left\{
\begin{array}{l l l}
 0 & \quad {\rm for} & \quad St <1, \\
\left.\frac{d i^2}{d t}\right|_{\rm i}+\left.\frac{d i^2}{d t}\right|_{\rm g}+
\left.\frac{d i^2}{d t}\right|_{\rm c}
 & \quad {\rm for} & \quad St \ge 1, 
\end{array}
\right.
\label{di2dt}
\end{eqnarray}
where the subscripts ``i'', ``g'', and ``c'' indicate the gravitational interaction, the gas drag, and the collisional damping.

The $e$ and $i$ evolution due to gravitational interaction is given by \citep{ohtsuki02}, 
\begin{eqnarray}
 \left. \frac{d e^2}{d t} \right|_{\rm i} &=& \Omega r^2 \int d m_2 n_{\rm s} (m_2) \frac{h_{1,2}^4 m_2}{(m_1+m_2)^2} 
\nonumber
\\
&& \times \left( m_2 P_{\rm VS} 
+ 0.7 P_{\rm DF}  \frac{ m_2 e^2 -m_1 e_1^2}{e_1^2+e_2^2}
\right), \\
\left. \frac{d i^2}{d t} \right|_{\rm i} 
&=& \Omega r^2 \int d m_2 n_{\rm s} (m_2) \frac{h_{1,2}^4 m_2}{(m_1+m_2)^2} 
\nonumber
\\
&& \times \left( m_2 Q_{\rm VS} 
+ 0.7 Q_{\rm DF}  \frac{ m_2 i^2 -m_1 i_1^2}{i_1^2+i_2^2}
\right), 
\end{eqnarray}
where 
\begin{eqnarray}
 P_{\rm VS} &=& 73 
\frac{\ln(10 \Lambda^2 /{\tilde e}_{1,2}^2+1)}{10 \Lambda^2/{\tilde e}_{1,2}^2}
+\frac{72 \Psi_{\rm PVS}(I) }{\pi {\tilde e}_{1,2} {\tilde i}_{1,2}} \ln(1+\lambda^2),
\\
 Q_{\rm VS} &=& (4 {\tilde i}^2+0.2 {\tilde e}^3 {\tilde i}) 
\frac{\ln(10 \Lambda^2 {\tilde e}_{1,2} + 1)}{10 \Lambda^2 {\tilde e}_{1,2}}
+\frac{72 \Psi_{\rm QVS}(I) }{\pi {\tilde e}_{1,2} {\tilde i}_{1,2}} \ln(1+\lambda^2), 
\\
 P_{\rm DF} &=& 10 {\tilde e}^2 
\frac{\ln (10 \Lambda^2 + 1)}{10 \Lambda^2}
+\frac{576 \Psi_{\rm PDF}(I) }{\pi {\tilde e}_{1,2} {\tilde i}_{1,2}} \ln(1+\lambda^2), \\
 Q_{\rm DF} &=& 10 {\tilde i}^2 
\frac{\ln (10 \Lambda^2 + 1)}{10 \Lambda^2}
+\frac{576 \Psi_{\rm QDF}(I) }{\pi {\tilde e}_{1,2} {\tilde i}_{1,2}} \ln(1+\lambda^2), 
\end{eqnarray}
with
\begin{eqnarray}
\Lambda &=& \frac{1}{3}({\tilde e}_{1,2}^2 +{\tilde i}_{1,2}^2 )
({\tilde i}_{1,2}^2 +1), \\ 
 \Psi_{\rm PVS}(I) &=& \int_0^1 \frac{5K\left(\frac{\sqrt{3(1-\lambda^2)}}{2}\right) 
- \frac{12(1-\lambda^2)}{1+3\lambda^2}E\left(\frac{\sqrt{3(1-\lambda^2)}}{2}\right) 
}{I+(I^{-1}-I)\lambda^2}d\lambda, \\
 \Psi_{\rm QVS}(I) &=& \int_0^1 \frac{K\left(\frac{\sqrt{3(1-\lambda^2)}}{2}\right) 
- \frac{12\lambda^2}{1+3\lambda^2}E\left(\frac{\sqrt{3(1-\lambda^2)}}{2}\right) 
}{I+(I^{-1}-I)\lambda^2}d\lambda, \\
 \Psi_{\rm PDF}(I) &=& \int_0^1 \frac{
\frac{1-\lambda^2}{1+3\lambda^2}E\left(\frac{\sqrt{3(1-\lambda^2)}}{2}\right) 
}{I+(I^{-1}-I)\lambda^2}d\lambda, \\
 \Psi_{\rm QDF}(I) &=& \int_0^1 \frac{
\frac{\lambda^2}{1+3\lambda^2}E\left(\frac{\sqrt{3(1-\lambda^2)}}{2}\right) 
}{I+(I^{-1}-I)\lambda^2}d\lambda.
\end{eqnarray}

Since $e$ and $i$ follow Rayleigh distributions, 
the evolution of mean $e$ and $i$ 
due to gas drag is given by \citep{adachi76,inaba01}
\begin{eqnarray}
 \left. \frac{d e^2}{d t} \right|_{\rm g} &=& -2 \frac{e^2 \Omega}{\eta St} 
\left( \frac{9 E(3/4)}{4 \pi} e^2 + \frac{i^2}{\pi} + \frac{9 \eta^2}{4}
\right)^{1/2}, \\
\left. \frac{d i^2}{d t} \right|_{\rm g} &=& -2 \frac{i^2 \Omega}{\eta St} 
\left( \frac{E(3/4)}{\pi} e^2 + \frac{4i^2}{\pi} + \eta^2
\right)^{1/2}. 
\end{eqnarray}
We only consider the leading-order terms of $e$ and $i$, 
because the higher-order terms of $e$ and $i$ are negligible for $e$ and $e$ with which we are concerned \citep{kobayashi15}. 

The collisional damping terms, $de^2/dt|_{\rm c}$ and $di^2/dt|_{\rm c}$ are given via the random velocities of collisional outcomes according to \citet{kobayashi+10}.

\section{relative velocity for strong coupling with gas
}
\label{sc:app_vrel}

To determine $\Delta v_{\rm rel,gas}$,
we use the vertical averaged values for $\Delta v_{\rm B}$, $\Delta v_r$, $\Delta v_\theta$, $\Delta v_z$, $\Delta v_{\rm t}$, given by 
\citep{adachi76, okuzumi12, ormel07}
\begin{eqnarray}
 \Delta v_{\rm B} &=& \sqrt{\frac{8 k_{\rm B} T(m_1+m_2)}{\pi m_1 m_2}},
  \\
 \Delta v_r &=& \left|\frac{St_1}{1+St_1^2}-\frac{St_2}{1+St_2^2}\right| 2\eta r \Omega, 
\\
 \Delta v_\theta &=& \left|\frac{1}{1+St_1^2}-\frac{1}{1+St_2^2}\right|\eta r \Omega, \\
 \Delta v_z &=& \left|\frac{St_1 }{1+St_1^2}-\frac{St_2}{1+St_2^2}\right|
\frac{h_{\rm s,1} h_{\rm s,2} \Omega}{h_{\rm s,1,2}},
\\
\Delta v_{\rm t} &=& (\Delta v_{\rm I}^2 + \Delta v_{\rm II}^2)^{1/2}, 
\end{eqnarray}
where %$T$ is the midplane temperature, $k_{\rm B}$ is the Boltzmann constant, 
\begin{eqnarray} 
\Delta v_{\rm I}^2 &=& \alpha_{\rm D} c_{\rm s}^2 \frac{St_1-St_2}{St_1+St_2} 
\left(\frac{St_1^2}{St_{1,2}^* + St_1}-\frac{St_1^2}{1 + St_1}
-\frac{St_2^2}{St_{1,2}^* + St_2}+\frac{St_2^2}{1 + St_2}
\right), 
\label{eq:v_I} 
\\
\Delta v_{\rm II}^2 &=& \alpha_{\rm D} c_{\rm s}^2 
(St_{1,2}^*-St_{\rm min})
\left(
\frac{(St_{1,2}^*+St_{\rm min})St_1+St_{1,2}^*St_{\rm min}
}{(St_1+St_{1,2}^*)(St_1+St_{\rm min})}
\right. 
\nonumber
\\&&\left. 
+
\frac{(St_{1,2}^*+St_{\rm min})St_2+St_{1,2}^*St_{\rm min}
}{(St_2+St_{1,2}^*)(St_2+St_{\rm min})}
\right),\label{eq:v_II} 
\end{eqnarray}
%$\rho_{\rm g}$ is the midplane gas density, $c_{\rm s}$ is the isothermal sound velocity, 
$St_{\rm min}^2 = 
\sqrt{\pi} \lambda_{\rm mfp} / 4 \sqrt{2} \alpha_{\rm D} h_{\rm g}$, and 
$St_{1,2}^* = {\rm MAX}(St_{\rm min},{\rm MIN}(1.6 St_1,1.))$ for $St_1
\geq St_2$.

\bibliography{ms}

\end{document}